\documentclass[english,twocolumn,english,aps,prx]{revtex4-1}
\usepackage[LGR,T1]{fontenc}
\usepackage[latin9]{inputenc}
\usepackage{amsmath}
\usepackage{amssymb}
\usepackage{graphicx}

\makeatletter

\ProvideTextCommand{\~}{LGR}[1]{\char126#1}
\usepackage{mathrsfs}

\makeatother

\usepackage{babel}
\begin{document}

\title{Quantum Sensing by Using STIRAP with Dressed States Driving}

\author{Hao Zhang$^{1,2}$, Guo-Qing Qin$^{1,2}$, Xue-Ke Song$^{3}$, and Gui-Lu Long$^{1,2,4,5}$}

\email{gllong@tsinghua.edu.cn}

\address{$^1$State Key Laboratory of Low-Dimensional Quantum Physics, Department of Physics, Tsinghua University, Beijing 100084, China\\
$^2$Frontier Science Center for Quantum Information, Beijing 100084, China\\
$^3$School of Physics and Material Science, Anhui University, Hefei 230601, China\\
$^4$Beijing National Research Center for Information Science and Technology, Beijing 100084, China\\
$^5$Beijing Academy of Quantum Information Sciences, Beijing 100193, China}

\date{\today}
\begin{abstract}
Exploring quantum technology to precisely measure physical quantities is a meaningful task for practical scientific researches.  Here, we propose a novel quantum sensing model based on dressed states driving (DSD) in stimulated Raman adiabatic passage. The model is universal for sensing different physical quantities, such as magnetic field, mass, rotation and etc. For different sensors, the used systems can range from macroscopic scale, e.g. optomechanical systems, to microscopic nanoscale, e.g. nitrogen-vacancy color centres in diamond. By investigating the dynamics of color detuning of DSD passage, the results show the sensitivity of sensors can be enhanced by tuning system with more adiabatic and accelerated processes in non-degenerate and degenerate color detuning regime, respectively. To show application examples, we apply our approach to build optomechanical mass sensor and solid spin magnetometer with practical parameters.\\
\end{abstract}

\maketitle

\section{Introduction}

Quantum sensing \cite{sensingrmp}, which uses the quantum property of a quantum system to build the sensor of a physical quantity, is a widely studied branch of quantum information and technology. 
According to the difference of physical quantities, such as magnetic field, temperature and mass, the required performance candidate has different physical systems known as solid spins \cite{NVPR,duan,lukinnature,Wrachtrupnature,Cappellaro,fwsunprapplied,dunc,zhaoNN,Panprl,LiuQE}, optomechanical systems \cite{MARMP2014,Dongscience,shen2016,liaoprl,YDWangPRL2012,LTianPRL2012,xiaopra,gong,Lvopto,Kuzykpra17,HZhangOE,XZhouLPL},  optical microcavity \cite{cavity1,cavity2,cavity3,cavity4,cavity5,cavity6,cavity7,cavity8} and etc. 
With the advantages of large range scale and high sensitivity, quantum sensing plays significant role in practical scientific researches and many interesting schemes have been proposed in recent years. For example, coupled with magnetic field, the nitrogen-vacancy (NV) color centres in diamond has been made for nanoscale magnetometer \cite{lukinnature,Wrachtrupnature,Cappellaro,fwsunprapplied,dunc} and  spins sensor  \cite{zhaoNN,Panprl}. Besides the solid spins, the sensors for detecting forces \cite{optoforce,qinsensing}, acceleration \cite{optoacce1,optoacce2} and masses \cite{optomass,Massprapplied} are proposed in optomechanical systems.

The perfect quantum operation is the goal pursued by quantum information tasks. For instance, quantum computing needs the fast quantum gate operations with high fidelity. To accurately control a quantum process, the adiabatic passage technique is proposed with good robustness \cite{NVVitanovARPC2001,KBergmannRMP1998,DGORMPSTA}. The widely used techniques for two-level and three-level system are called rapid adiabatic passage \cite{NVVitanovARPC2001} and stimulated Raman adiabatic passage (STIRAP) \cite{KBergmannRMP1998}, respectively. However, the trade off between the robustness and speed limits the potential of adiabatic passage technique used for fast quantum operation. To solve the above problem, the shortcut to adiabaticity is developed \cite{DGORMPSTA,MVBerry2009,XChenPRL2010,ABaksicPRL2016,YHChenpra,
YLiangPRA2015,XKSongNJP2016} and applied successfully in experiment in different systems, including Bose-Einstein condensates in optical lattices \cite{MGBason2012}, cold atoms \cite{YXDuNC2016,OLguo}, trapped ions \cite{AnNC2016} and NV centers in diamond \cite{JZhangprl2013,BBZhou2016,PRLNV2019}.
The direct way, so called transitionless driving \cite{MVBerry2009,XChenPRL2010}, to realize the shortcut to adiabaticity is to suppress or cancel the undesired transition terms of Hamiltonian in adiabatic frame and this approach is generalized and improved for realizable in situation without constructing the forbidden transition in three-level system by using dressed states driving (DSD) \cite{ABaksicPRL2016,BBZhou2016}.

In this paper, we propose an universal quantum sensing model based on DSD in stimulated Raman adiabatic passage. By investigating the detuning dynamic of three-level system governed by DSD, we find that the population of target state has different properties of sensitivity in degenerate or non-degenerate color detuning area. As the evolution speed becomes faster, the slope of variation of the population becomes larger and smaller for degenerate or non-degenerate color detuning, respectively. According to this property, different kinds of quantum sensors can be built by constructing the relationship between detuning and corresponding physical quantities. We show the examples for realizing the improved sensitivity optomechanical mass sensor and solid spin magnetometer based on degenerate and non-degenerate color regime, respectively.

This article is organized as follows:  We first introduce our universal model for quantum sensing in three-level system in Sec.~\ref{sec2}. In Sec.~\ref{sec3}, we calculate basic physics about the detuning dynamics of DSD passage. To show practical applications, we perform how to use our model to realized the mass sensor and magnetometer in Sec.~\ref{sec41} and Sec.~\ref{sec42}, respectively. A conclusion is given in Sec.~\ref{sec5}.

\begin{figure}[!ht]
\begin{center}
\includegraphics[width=8.0cm,angle=0]{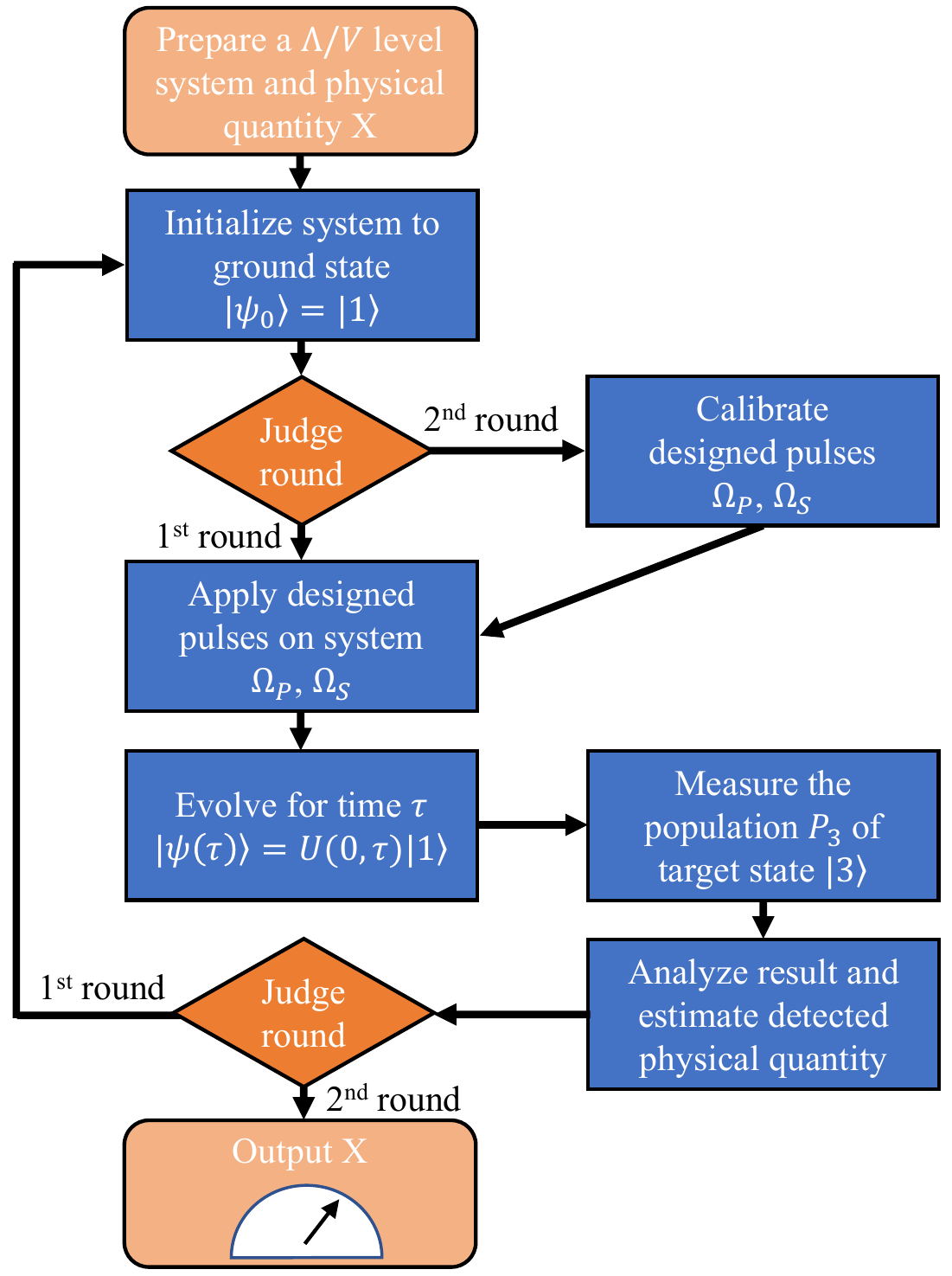}
\caption{Flowchart for the quantum sensing in three level system. $X$ is a unknown physical quantity to be measured.}\label{flowchart}
\end{center}
\end{figure}

\section{Universal model for quantum sensing in three-level system} \label{sec2}
To realize quantum sensing for unknown physical quantity, labelled with X, the first task is to find a controllable quantum system which can couple with corresponding X well. For a given initial state, the evolution paths of the system are different with different values of X. The key point is one should theoretically give a relationship between observable quantity Y of the system and X, the form can be written as
\begin{eqnarray}      \label{yfx}
Y=f(X).
\end{eqnarray}
To detect the magnitude of Y, one can infer the quantity X via the relationship Eq.(\ref{yfx}). The sensitivity of sensor can be enhanced  by improving the absolute value of differential coefficient $\frac{\partial Y}{\partial X}$. 

Here we consider an universal three-level (mode) system and the detailed flowchart of our sensing proposal is given in Fig.~\ref{flowchart}. 
The first step is to find a three-level (mode) system which can couple with X. Before evolution, the system should be prepared on its initial state $|1\rangle$, e.g. the ground state of the system. This step is so-called initialization. If the operation is the first round, we apply two pulses designed under the zero physical quantity X. After a time $\tau$, the system evolves to the final state and we measure the population of the separated level $|3\rangle$. The pulses used here are calculated for perfect population transfer between states $|1\rangle$ and $|3\rangle$. If the sensing has low resolution, the pulses should be calibrated to promise the sensor in high sensitivity domain. In this situation shown in Fig.~\ref{flowchart}, the sensing process has the second round.

\begin{figure}[ht]
\begin{center}
\includegraphics[width=8.3 cm,angle=0]{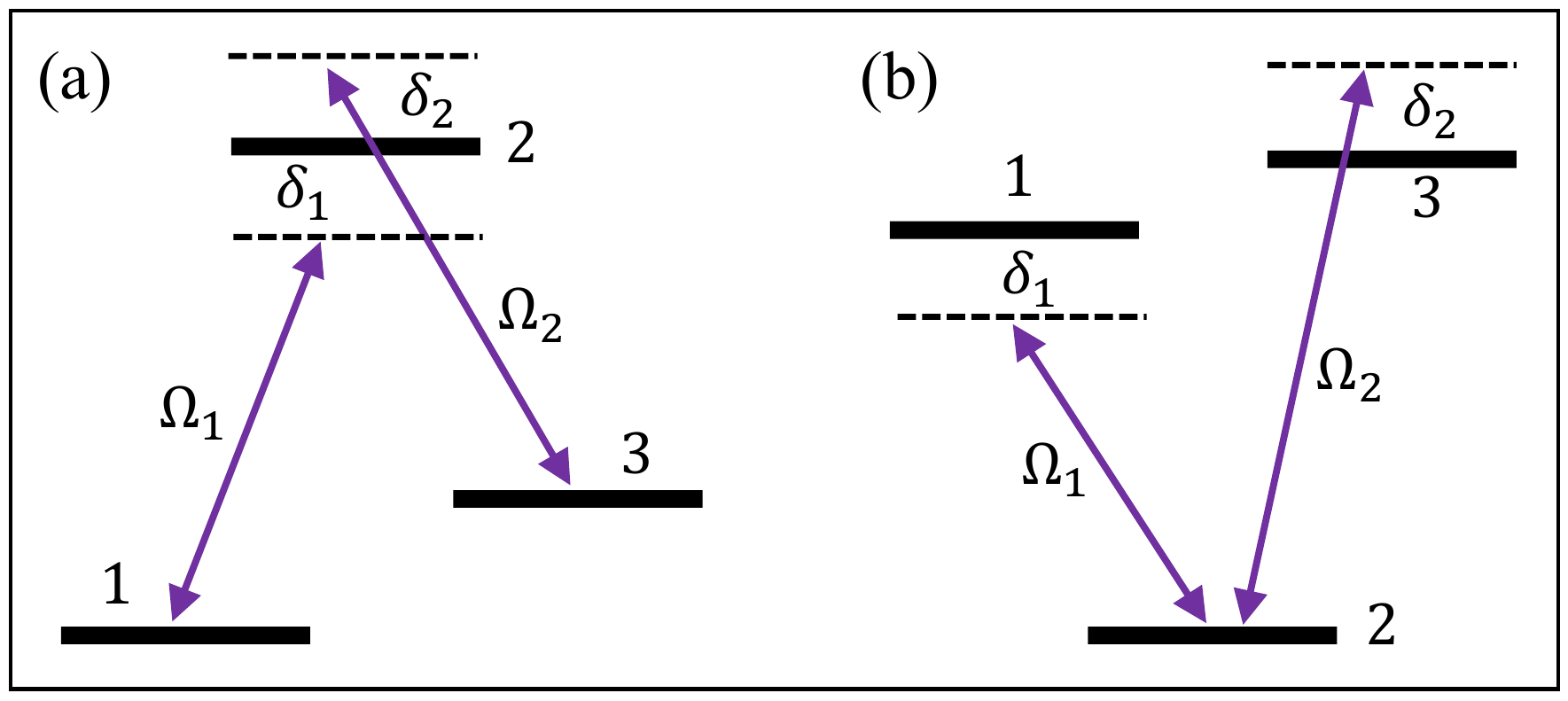}
\caption{Schematic of generic three-level system. (a) $\Lambda$ type level structure with detuning. (b) $V$ type level structure with detuning.}\label{type}
\end{center}
\end{figure}


\section{Color detuning dynamics of DSD passage}\label{sec3}
We consider two different kinds of structure of generic three-level systems, i.e. $\Lambda$ and $V$, shown in Fig.~\ref{type}. With driving pulses $\Omega_{1}$ and $\Omega_{2}$, two type structures have the Hamiltonian in the rotating frame given by 
\begin{eqnarray}      \label{eqM1}
H_{1}(t)=
\left[
\begin{array}{ccc}
\delta_{1}&\Omega_{1}(t)&0\\
\Omega_{1}(t)&0&\Omega_{2}(t)\\
0&\Omega_{2}(t)&\delta_{2}\\
\end{array}
\right].
\end{eqnarray}
For multimode bosonic system, the matrix $H_{1}(t)$ satisfies equation $id\vec{v}(t)/dt=H_{1}(t)\vec{v}(t)$, where the vector operator is $\vec{v}(t)=[a_{1}(t),a_{2}(t),a_{3}(t)]^{T}$. The $a_{i}$ $(a^{\dag}_{i})$ $(i=1,2,3)$ is defined as the annihilation (creation) operators for the corresponding $i$-th mode with the frequency $\omega_{i}$, respectively. In fermi system, the $\omega_{i}$ is energy of $i$-th  bare level.  $\delta_i$ is  laser detuning with corresponding levels (modes). $\Omega_{i}(t)$ is the effective coupling strength between the corresponding levels (modes) shown in Fig.~\ref{type}.   When all the detuning are zero, i.e. $\delta_{i}=0$, the system has the eigenvalues $\lambda=0,\pm \Omega_{0}$ and the corresponding eigenstates (eigenmodes) are dark state (mode) $\psi_{d}=[-\Omega_{2}/\Omega_{0},0,\Omega_{1}/\Omega_{0}]^{T}$ and bright state (mode) $\psi_{\pm}=[\Omega_{1}/\Omega_{0},\pm1,\Omega_{2}/\Omega_{0}]^{T}/\sqrt{2}$, where $\Omega_{0}=\sqrt{\Omega^{2}_{1}+\Omega^{2}_{2}}$.


\begin{figure*}[!ht]
\begin{center}
\includegraphics[width=17.5 cm,angle=0]{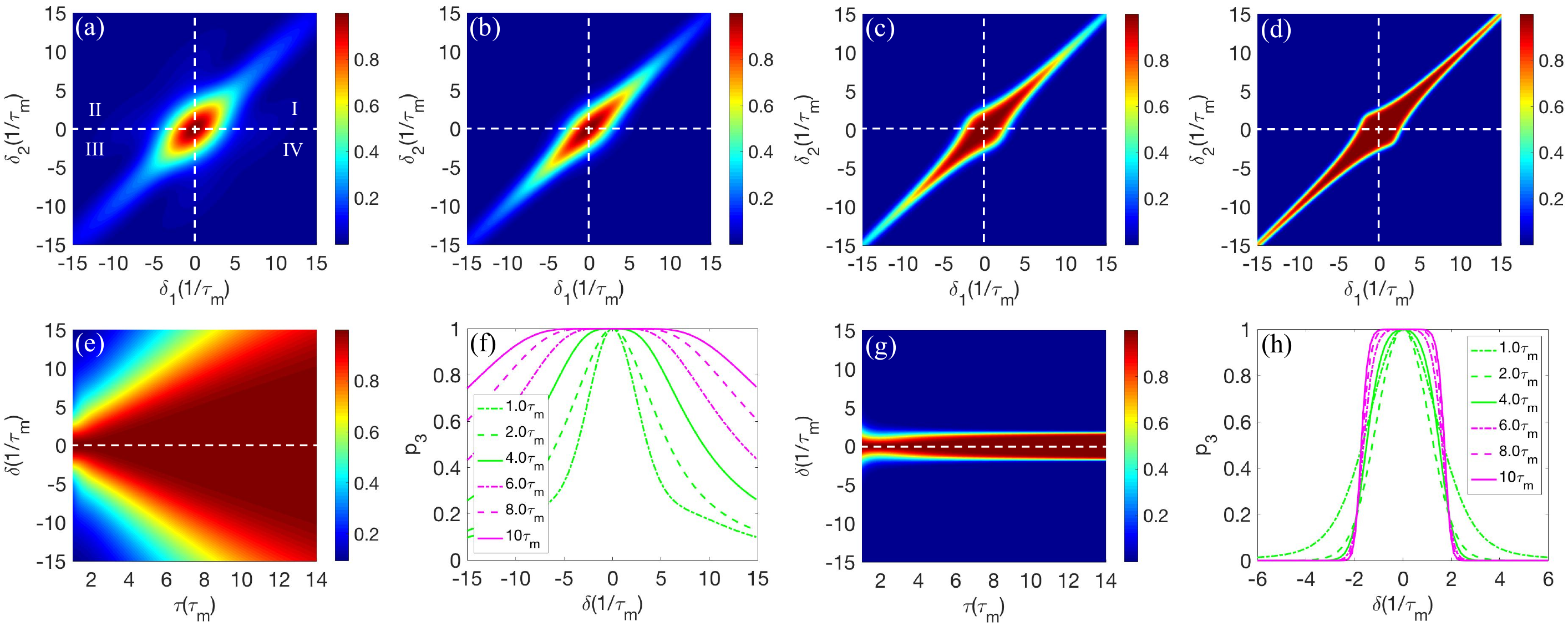}
\caption{Dynamics of three-level system with different detuning. (a)-(d) Population of level 3 vs detuning $\delta_1$ and $\delta_2$ with different pulse duration. (a) $\tau=1\tau_m$; (b) $\tau=2\tau_m$; (c) $\tau=5\tau_m$; (d) $\tau=10\tau_m$; (e) Population of level 3 vs  same color detuning $\delta=\delta_1=\delta_2$ and pulse duration $\tau$. (f) Curves of some particular duration times for same color detuning. (g) Population of level 3 vs  different color detuning $\delta=\delta_1=-\delta_2$ and pulse duration $\tau$. (h) Curves of some particular duration times for different color detuning.}\label{dynamics}
\end{center}
\end{figure*}

To construct an STIRAP, we choose the adiabatic pulses with `Vitanov' envelope given by \cite{Vasilevpra}
\begin{eqnarray}        \label{eqVitanov}
&\Omega_{1}(t)=\Omega_{0}\sin[\theta(t)],\nonumber\\
&\Omega_{2}(t)=\Omega_{0}\cos[\theta(t)],\nonumber\\
&\theta(t)=\frac{\pi}{2}\frac{1}{1+e^{-t/\tau}},
\end{eqnarray}
where $\tau$ is pulse duration. We assume that the initial state is prepared in level 1 and the target state is level 3. The population can be transferred successfully via STIRAP. To speed up the STIRAP, the DSD method modulates the pulse envelope with \cite{ABaksicPRL2016}
\begin{eqnarray}        \label{DSD}
&\widetilde{\theta}(t)=\theta(t)-\arctan(\dfrac{g_{x}(t)}{\Omega_{0}(t)+g_{z}(t)}),\nonumber\\
&\widetilde{\Omega}_{0}(t)=\sqrt{(\Omega_{0}(t)+g_{z}(t))^{2}+g_{x}^{2}(t)},
\end{eqnarray}
where $g_{x}(t)$, $g_{z}(t)$ and $\mu$ are chosen with $g_{x}(t)=\dot{\mu}$, $g_{z}(t)=0$,  $\mu=-\arctan(\dfrac{\dot{\theta}(t)}{\Omega(t)})$, respectively. As shown in Fig.~\ref{dynamics}, the population of level 3 is changed with respect to the detuning $\delta_1$ and $\delta_2$. From Fig.~\ref{dynamics} (a) to (d), the pulse durations are chosen with $\tau=1\tau_m$, $2\tau_m$, $5\tau_m$ and $10\tau_m$, respectively. The $\tau_m$ is the minimum pulse duration time and has the relationship $\tau_m\simeq1/2.63\Omega_0$ \cite{ABaksicPRL2016}. In degenerate color detuning area, i.e. $\delta_1=\delta_2$, the bright color area becomes larger, it means that the population becomes less sensitive to the detuning as the evolution time is longer. When the duration time is in its limitation, i.e. $\tau=1.0\tau_m$, the variation of population is limited in the smallest area. On the contrary, the bright color area which becomes smaller in II and IV area in the coordinate frame indicates that the variation of population with non-degenerate color detuning, i.e. $\delta_1=-\delta_2$, is more sensitive to the detuning, when the duration time becomes longer. To show the results clearly, we plot the population with respect to the detuning and duration time with degenerate and non-degenerate color detuning in Fig.~\ref{dynamics} (e) and (g), respectively. Some curves with particular duration time are shown in Fig.~\ref{dynamics} (f) and (h) for degenerate and non-degenerate color detuning, respectively. In Fig.~\ref{dynamics} (f), as the duration time becomes shorter, the slope of the curve becomes steeper and the curve of $\tau=1.0\tau_m$ is the steepest one. However, in Fig.~\ref{dynamics} (h), the variation of population from 1 to 0 is faster in detuning scale as the duration time becomes longer. The results indicates that the accelerated passage enhances the sensitivity of population variation in degenerate color detuning regime, but for non-degenerate color regime, the sensitivity is enhanced by more adiabatic passage.

\section{Applications for sensing}\label{sec4}

\subsection{Degenerate color optomechanical mass sensor}\label{sec41}

\begin{figure}[!ht]
\begin{center}
\includegraphics[width=8.5cm,angle=0]{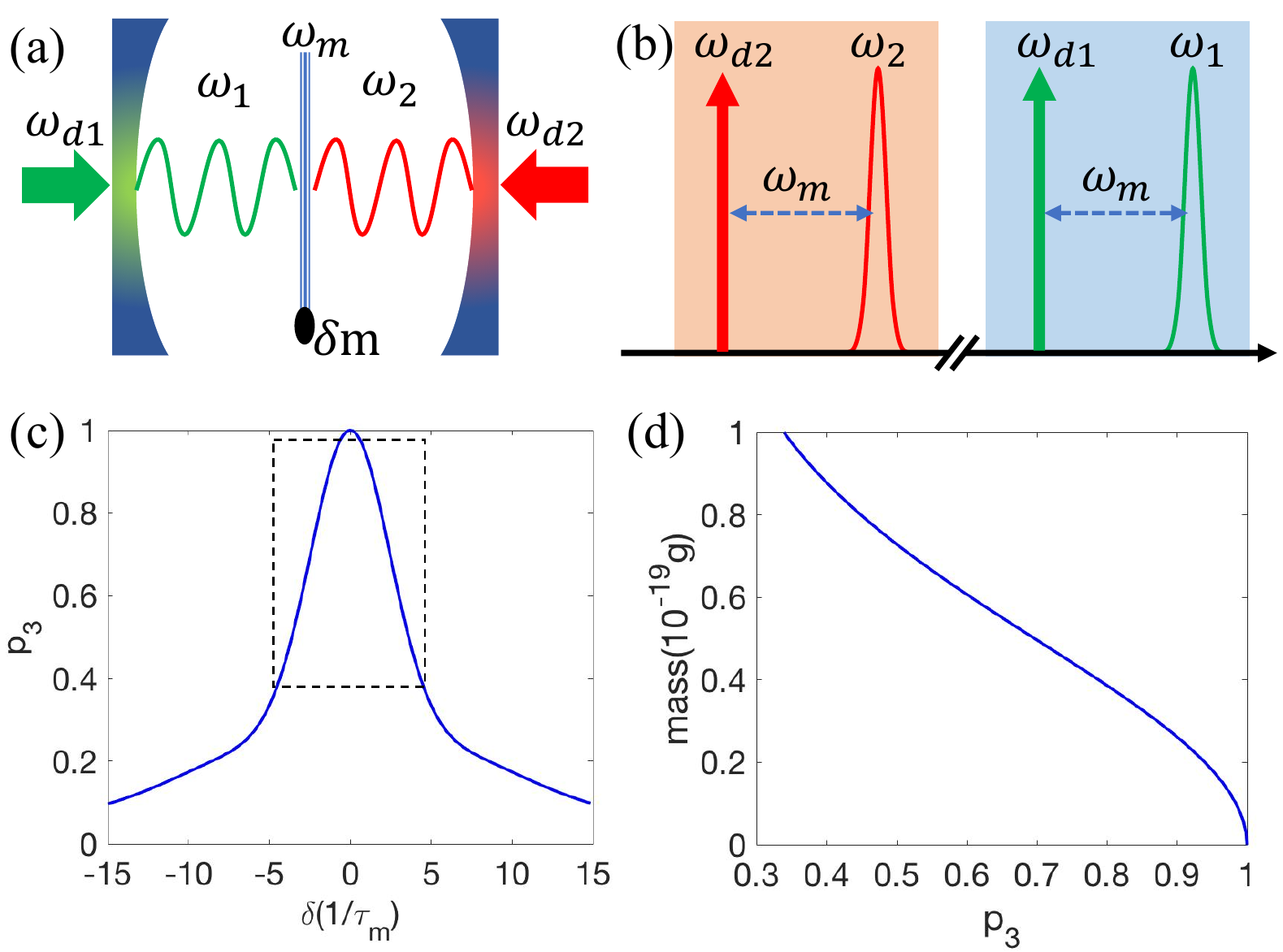}
\caption{Schematic of same color optomechanical mass sensor. (a) Generic setup for optomechanical system. Two optical cavities couple to each other by driving a mechanical oscillator. The small dot represents a deposited mass on the mechanical oscillator. (b) Red detuning drivings for two optomechanical interactions. (c) The curve for population vs detuning with duration $\tau=1.0\tau_{m}$. (d) The mass sensor in the domain labelled with dashed rectangle in (c).}\label{mass}
\end{center}
\end{figure}

To build a mass sensor, we consider a multimode optomechanical model composed of two optical cavities and a mechanical membrane. The detailed setup is shown in Fig.~\ref{mass}(a). By pumping two laser fields with frequency $\omega_{d1}$ and $\omega_{d2}$, two optical cavities couple to middle mechanical oscillator simultaneously. The frequency of left, right cavity and mechanical oscillator are $\omega_{1}$, $\omega_{2}$ and $\omega_{m}$, respectively. The black small dot in mechanical membrane is deposited mass with magnitude $\delta m$ to be measured. After the standard linearization procedure, the Hamiltonian of this multimode interactions structure is given by \cite{LTianPRL2012,YDWangPRL2012,xiaopra,Kuzykpra17,HZhangOE,XZhouLPL}
\begin{eqnarray}        \label{eqmassH1}
H_{op1}\!=\omega_{m}b^{\dag}b\!+\!\!\!\sum_{i=1,2} \!\!\left[-\Delta_{i}a^{\dag}_{i}a_{i}\!+\!G_{i}(a^{\dag}_{i}\!+\!a_{i})(b\!+\!b^{\dag})\right]\!\!,\;\;
\end{eqnarray}
where $a_{i}$ $(a^{\dag}_{i})$ $(i=1,2)$ and $b$ $(b^{\dag})$ are the annihilation (creation) operators for the $i$-th cavity mode and the mechanical mode, respectively. Here, $\Delta_{i}=\omega_{di}-\omega_{i}$ and $G_{i}=G_{0i}\sqrt{n_{i}}$ are the pump laser detuning with cavity mode and the effective optomechanical coupling strength, respectively. $G_{0i}$ and $n_{i}$ are the single-photon optomechanical coupling and the intra-cavity photon number produced by the driving field, respectively. When we consider the situation that all the cavities are driven with red detuning laser shown in Fig.~\ref{mass} (b), under the rotating-wave approximation, the Hamiltonian can be rewritten as 
\begin{eqnarray}      \label{eqmassH2}
H_{op2}=\sum_{i=1,2}\delta_{i}a_{i}^{\dag}a_{i}+G_{i}(a_{i}^{\dag}b+b^{\dag}a_{i}).
\end{eqnarray}
Here, the detuning is $\delta_{i}=-\Delta_{i}-\omega_{m}$. Calculating the  Heisenberg equations of above Hamiltonian, the effective coefficient matrix is equivalent to the Eq. (\ref{eqM1}) by defining the vector operator with $\vec{v}_{op}(t)=[a_{1}(t),b(t),a_{2}(t)]^{T}$. The coupling strength $G_i$ is chosen with the form in Eq. (\ref{DSD}) to realized the population transfer from cavity 1 to 2.  To realize the mass sensor, we first assume that the mechanical membrane is deposited a mass represented with a black dot shown in Fig.~\ref{mass}(a). Therefore, the frequency of mechanical mode has a little drift and the resonance coupling is broken by inducing the two little detuning with degenerate color. This additional mass impacts the result of the population in cavity 2. The relationship between frequency shift $\delta\omega_m$ and the deposited mass $\delta m$ is given by \cite{Massprapplied} 
\begin{eqnarray}      \label{eqmassshift}
\delta\omega_m=R\delta m.
\end{eqnarray}
The parameter $R=\omega_m/2m$ is mass responsivity, where $m$ is the mass of the mechanical resonator for supporting the deposition. By choosing the detuning with $\delta_1=\delta_2=0$ without deposited mass, when we assume that the mechanical resonator is added with a deposited mass $\delta m$, the system evolves under the degenerate color detuning $\delta_1=\delta_2=-R\delta m$. Therefore, the variation of population of cavity 2 infers the magnitude of $\delta m$. According to the Fig.~\ref{dynamics} (e) and (f), one can choose the fastest evolution path, i.e. duration $\tau=1.0\tau_m$, to design the mass sensor with high sensitivity and consider the domain in the rectangle in Fig \ref{mass}(c). To show the practical example, the curve of sensing deposited mass of mechanical oscillator is performed in Fig \ref{mass}(d) with given parameters given as $\omega_m=2\pi\times6$ GHz and $m=10^{-15}$ g. The $G_0$ which is equal to the $\Omega_0$ of Eq. (\ref{eqVitanov}) is assumed with $G_0\sim 1.0$ MHz.  The results show if the resolution of population is about $0.1$, the optomechanical mass sensor has the resolution with scale $10^{-20}$ g for a mechanical oscillator with mass $m=10^{-15}$ g.

\subsection{Non-degenerate color solid spin magnetometer}\label{sec42}
\begin{figure}[!ht]
\begin{center}
\includegraphics[width=8.2cm,angle=0]{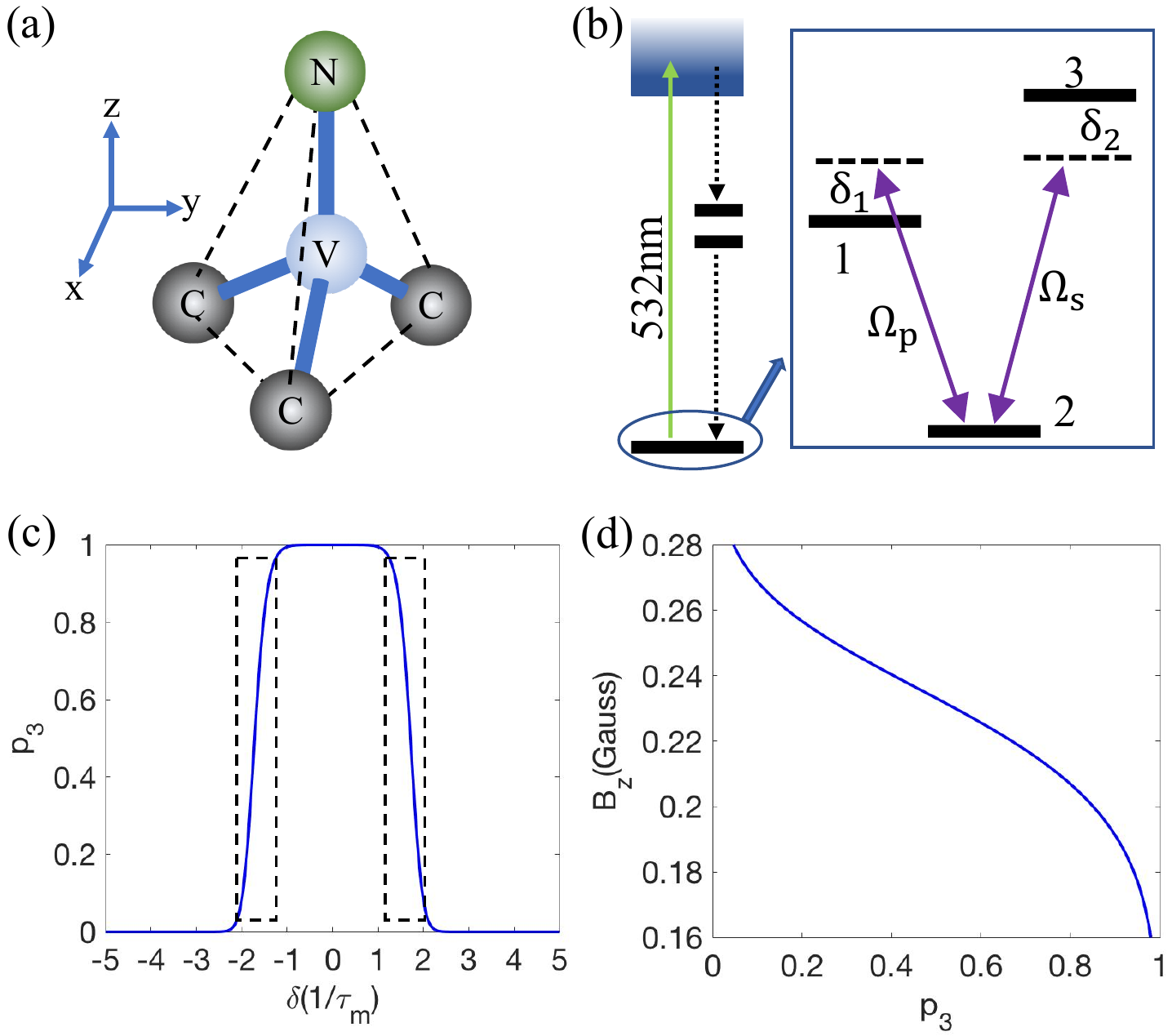}
\caption{Schematic of solid spin for detecting magnetic field. (a) The structure of an NV color centres in diamond. (b) Level structure of NV centre in diamond. The levels in the right rectangle are triplet structure in ground state. (c) The curve for population vs detuning with duration $\tau=10\tau_{m}$. (d) The magnetometer  in the domain labelled with dashed rectangle in (c).}\label{NVS}
\end{center}
\end{figure}

As a solid spin system,  NV centres in diamond is a promising platform for designing magnetometer due to its good coupling between spin and external magnetic field. As shown in Fig.~\ref{NVS}, the NV centres have a rich level structure. In the environment with a weak static magnetic field along the NV principle axis, the Hamiltonian of electron-spin ground triplet is given by \cite{NVPR}
\begin{eqnarray}        \label{NVH}
H_{NV}=DS_{z}^{2}+\gamma_{e}B_{z}S_{z},
\end{eqnarray}
where $D=2.87\ \text{GHz}$ is zero-field splitting between spin triplet $m_{s}=0$ and $m_{s}=\pm1$. $S_{z}$ is three dimension electron-spin operator. $\gamma_{e}=2.8025\ \text{MHz}\cdot\text{Gauss}^{-1}$  is the electronic gyromagnetic ratio.  The first term describes the zero-field splitting and the second one is electronic-spin Zeeman splitting term. 
The NV centres in diamond with the Hamiltonian as Eq. (\ref{NVH}) has the level structure shown in the right rectangle of Fig.~\ref{NVS} (b). The splitting of levels in $m_{s}=\pm1$ induced by magnetic field produces non-degenerate color detuning, as the detuning in Fig.~\ref{NVS} (b) has relationship with $\delta=-\delta_{1}=\delta_{2}=\gamma_{e}B_{z}$. Therefore, by applying two microwave pulses in the non-degenerate color regime, the population of level 3 of NV centre infers the magnitude of the magnetic field via the electronic gyromagnetic ratio.


To build the magnetometer, we design the driving pulses with the parameters $\tau=10\tau_m$. The interaction domain of detuning is chosen in the rectangle shown in Fig.~\ref{NVS} (c). Therefore, if the magnetic field is so weak, the frequency of pulse should be calibrated in effective detuning domain, i.e. in the dashed rectangle in Fig.~\ref{NVS} (c), before formal measurement, as shown in the first round of flowchart Fig.~\ref{flowchart}. When the parameters are given with $\Omega_0\sim 1.0$ MHz, the sensing curve of magnetic field is shown in Fig.~\ref{NVS} (d). The scale of the magnetometer is about $10^{-2}\ \text{Gauss}$.

\section{CONCLUSION}\label{sec5}

To conclude, we use the three-level (mode) system to design an universal quantum sensing model by using DSD. Driving with different pulses, three-level system has two different interaction regime, i.e. degenerate and non-degenerate color detuning. In degenerate color detuning, the variation of population in target state is more sensitive to the detuning value as the process is accelerated. On the contrary in non-degenerate color detuning, the process is more adiabatic, the sensitivity of population of target state is higher. Those properties indicate the system can be used to design different kinds of quantum sensors with higher sensitivity by chosen with different interaction regime. To show the examples of practical sensor, we perform how to apply our model for building degenerate color mass sensor and non-degenerate color magnetometer.

\begin{acknowledgments}
This work was supported by the China Postdoctoral Science Foundation under Grant No.2019M650620; National Natural Science Foundation of China under Grants (61727801); National Key Research and Development Program of China (2017YFA0303700); Beijing Advanced Innovation Center for Future Chip (ICFC); The Key Research and Development Program of Guangdong province (2018B030325002).  
\end{acknowledgments}


\end{document}